\documentclass[aps,pre,superscriptaddress,twocolumn,showpacs]{revtex4}
\usepackage{epsfig}

\newcommand{\be}{\begin{equation}}
\newcommand{\ee}{\end{equation}}
\newcommand{\bc}{\begin{center}}
\newcommand{\ec}{\end{center}}
\newcommand{\bi}{\begin{itemize}}
\newcommand{\ei}{\end{itemize}}
\newcommand{\ba}{\begin{eqnarray}}
\newcommand{\ea}{\end{eqnarray}}

\newcommand{\ignore}[1]{}

\begin{document}

\title{Cascade Dynamics of Multiplex Propagation}

\author{Damon Centola}
\email{dc288@cornell.edu} \affiliation{Department of Sociology,
Columbia University, New York, NY 10027} \affiliation{Department
of Sociology, Cornell University, Ithaca, NY 14853}
\homepage{http://hsd.soc.cornell.edu}
\author{Michael W. Macy}
\affiliation{Department of Sociology, Cornell University, Ithaca,
NY 14853}
\homepage{http://hsd.soc.cornell.edu}
\author{V\'{\i}ctor M. Egu\'{\i}luz}
\email{victor@imedea.uib.es}
\homepage{http://www.imedea.uib.es/physdept}
\affiliation{Instituto Mediterr\'aneo de Estudios Avanzados IMEDEA
(CSIC-UIB), E-07122 Palma de Mallorca, Spain}

\date{\today}

\begin{abstract}

Random links between otherwise distant nodes can greatly
facilitate the propagation of disease or information, provided
contagion can be transmitted by a single active node. However we
show that when the propagation requires simultaneous exposure to
multiple sources of activation, called {\em multiplex
propagation}, the effect of random links is just the opposite: it
makes the propagation more difficult to achieve. We calculate
analytical and numerically critical points for a threshold model
in several classes of complex networks, including an empirical
social network.

\end{abstract}

\maketitle

%%%%%%%%%%%%%%%%%%%%%%%%%%%%%%%%%%%%%%%%%%%%%%%%%%%%%%%%%%%%%%%%%%%%%%%%%%%%%%%
{\em Introduction.}-- Recently much attention has been paid to
complex networks as the skeleton of complex systems
\cite{Watts98,Strogatz01,Albert02,Dorogovtsev02,Newman03}. For
example, recent advances in complex systems have shown that most
real networks display the small world property: they are as
clustered as a regular lattice but with an average path length
similar to a random network \cite{Watts98}. More precisely, it has
been shown that surprisingly few bridge links are needed to give
even highly clustered networks the ``degrees of separation"
characteristic of a ``small world". Interestingly, these random
links significantly increase the rate of propagation of contagions
such as disease and information \cite{Watts98,Watts99,Newman00}.
For {\em simple propagation} --such as the spread of information
or disease-- in which a single active node is sufficient to
trigger the activation of its neighbors, random links connecting
otherwise distant nodes achieve dramatic gains in propagation
rates by creating ``shortcuts" across the graph
\cite{Granovetter73,Morris00}. Sociologists have long argued that
bridge links between disjoint neighborhoods promote the diffusion
of information and innovation, a regularity known as the
``strength of weak links" \cite{Granovetter73}.

In addition to simple propagation, {\em multiplex propagation}, in
which node activation requires simultaneous exposure to multiple
active neighbors, is also common in the social world. Fads, stock
market herds, lynch mobs, riots, grass roots movements, and
environmental campaigns (such as curb side recycling) share the
important property that a bystander's probability of joining
increases with the level of local participation by her neighbors
\cite{Granovetter78}. In this case, one neighbor acting alone is
rarely sufficient to trigger a cascade. These cascades often
display a second important property: they typically unfold in
clustered networks. Empirical studies have consistently found that
recruitment to social movements is most effective in locally dense
networks characterized by strong interpersonal ties
\cite{McAdam86,McAdam93}. Short cycles expose potential recruits
to multiple and overlapping influences that provide the strong
social support required to motivate costly investments of time,
effort, and resources.

In this paper, we use a threshold model to analyze the effect of
bridge ties in complex networks on the dynamics of multiplex
propagation. \cite{Granovetter78,Watts02,Dodds04}. Our results
show that contrary to the results for cascades in small worlds
networks \cite{Watts98,Watts99,Newman00}, for multiplex
propagation, random links to distant nodes reduce propagation
rates. Moreover, too many random links can prevent cascades from
occurring altogether. To test the results of our model, we examine
its predictions on an empirical network with scale-free degree
distribution.

%%%%%%%%%%%%%%%%%%%%%%%%%%%%%%%%%%%%%%%%%%%%%%%%%%%%%%%%%%%%%%%%%%%%%%%%%%%%
{\em The threshold model.}-- The system is composed of a set of
$N$ agents located at the nodes of a network. Each agent can be in
one of two states: $1$ indicates that the agent is active,
otherwise its state is $0$. Each agent is characterized by a fixed
threshold $0\le T\le 1$. The dynamics are defined as follows. Each
time $t$ a node $i$ is selected at random. Then
\begin{enumerate}
\item if its state is 1 (active), then it will remain active;
\item however, if its state is 0, then it becomes active, changing
its state to $1$, if only if the fraction of its neighbors in the
active state is equal to or larger than $T$.
\end{enumerate}

In order to isolate the effect of the network topology from the
effect of threshold distribution, we assign every node an
identical threshold $T$, which determines the fraction of
neighbors required to activate it. By definition, a single active
seed is insufficient for multiplex propagation. Hence, we seed the
network by randomly selecting a focal node and activating this
node and all of its neighbors. For any graph, there is a critical
threshold $T_c$ above which propagation is not possible.

%%%%%%%%%%%%%%%%%%%%%%%%%%%%%%%%%%%%%%%%%%%%%%%%%%%%%%%%%%%%%%%%
{\em Critical Thresholds in Regular and Random Graphs.}-- First,
we compare critical thresholds in random and regular networks with
identical size $N$ and average degree $\langle k \rangle$.

%%%%%%%%%%%%%%%%%%%%%%%%%%%%%%%%%%%%%%%%%%%%%%%%%%%%%%%%%%%%%%%Fig. 1
\begin{figure}
\epsfig{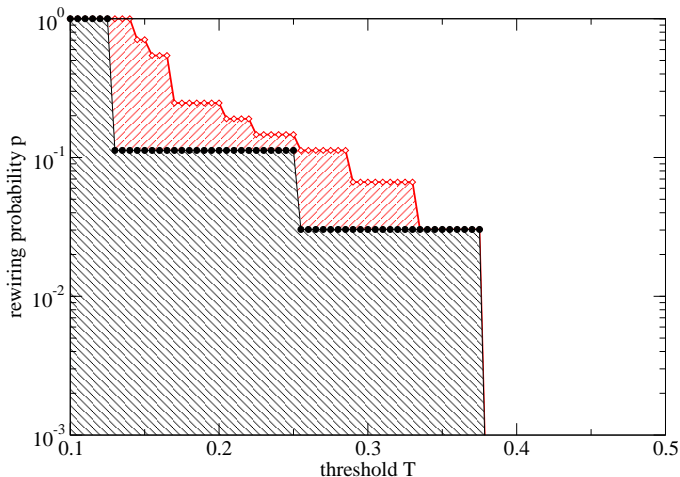}
\caption{
Cascade window for small-world networks. Results are averaged over
1000 realizations in a 10,000 node network. The shaded area
indicates when cascades occur for small-world networks obtained
after rewiring (red diamonds) or permuting (black circles) links.
}
\label{fig1}
\end{figure}

As defined by the cascade condition in Ref.~\cite{Watts02}, for a
random graph of size $N$ in which all the nodes have the same
degree $\langle k \rangle \ll N$ and the same threshold $T$, as
$N$ approaches infinity the probability that two nodes in the
initial seed neighborhood will have a common neighbor approaches
zero. Thus, the critical threshold for a random graph is
approximated by \be T_c^r = \frac{1}{\langle k \rangle}~,
\label{tcrand} \ee which corresponds to the limiting case of
simple propagation, and shows that multiplex propagation cannot
succeed on sparse random graphs.

The critical threshold for a regular one-dimensional lattice is
\cite{Morris00} \be T_c^{1d} = \frac{1}{2}~. \label{tc1d} \ee
While in a one-dimensional ring with average degree $\langle k
\rangle$ the critical threshold is independent of the interaction
length [Eq.~(\ref{tc1d})], in a random graph with the same average
degree $\langle k \rangle$ the critical threshold decreases with
$\langle k \rangle$ [Eq.~(\ref{tcrand})]. Thus, the difference
between the critical thresholds of regular one-dimensional
lattices and random networks increases with the average degree
$\langle k \rangle$, making the one-dimensional lattice much more
vulnerable to multiplex propagation than an equivalent random
network.

This feature is also observed in two-dimensional lattices. In a
two-dimensional lattice with near and next-nearest neighbors (also
called a Moore neighborhood) the critical threshold is
\cite{Morris00} \be T_c^{2dnn}=\frac{3}{8}=0.375~. \label{tc2dM}
\ee

As the interaction length in the two-dimensional lattice
increases, the critical threshold approaches the upper limit of
$1/2$ \cite{Morris00}. Thus, increasing $\langle k \rangle$
increases the differences in the critical thresholds between
regular and random networks, making clustered regular networks
able to support comparatively greater amounts of multiplex
propagation than random networks.

%%%%%%%%%%%%%%%%%%%%%%%%%%%%%%%%%%%%%%%%%%%%%%%%%%%%%%%%%%%%%%%Fig. 2
\begin{figure}
\epsfig{file=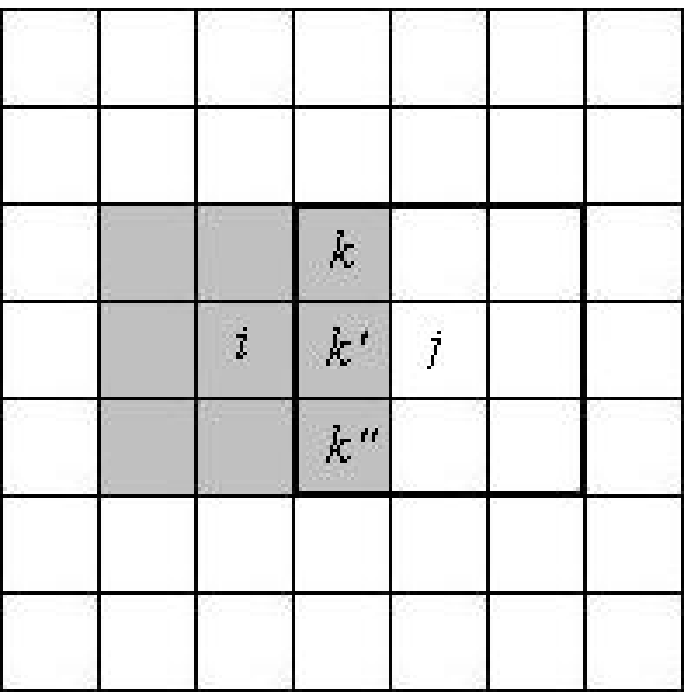,width=0.2\textwidth,angle=0}
\epsfig{file=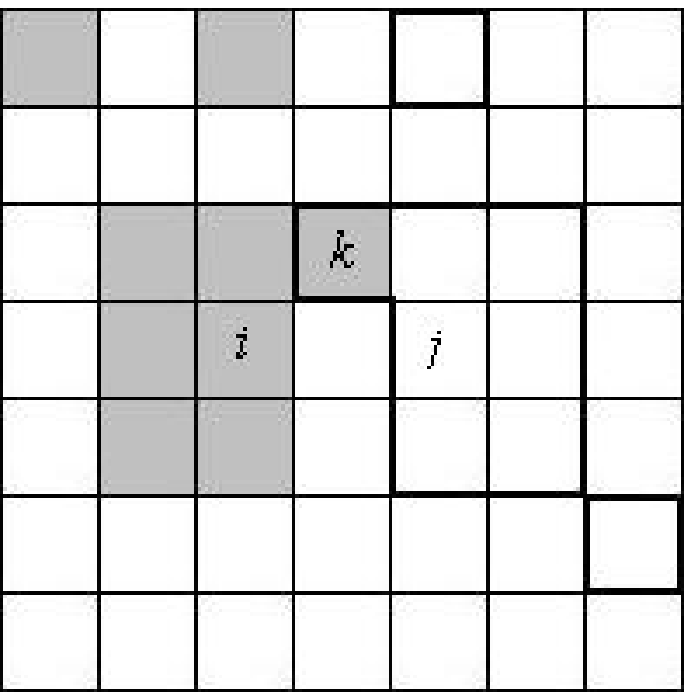,width=0.2\textwidth,angle=0}
\caption{Bridges between neighborhoods in (left) a regular lattice
and (right) after some rewiring. Rewiring deceases the bridge
width between $i$ and $j$ making cascade propagation more
difficult.}
\label{fig2}
\end{figure}

%%%%%%%%%%%%%%%%%%%%%%%%%%%%%%%%%%%%%%%%%%%%%%%%%%%%%%%%%%%%%%%%%%%%%%%%%%%%%%
{\em Small-world networks.}-- We next explore the transition in
critical thresholds that occurs in the small-world regime between
perfect regularity and pure randomness by randomizing links in a
two-dimensional regular lattice with nearest and next-nearest
neighbors. We study the effects of bridge ties on the success of
multiplex propagation using two different perturbation algorithms.
One is the usual {\em rewiring} technique \cite{Watts98}: each
link is broken with probability $p$ and reconnected to a randomly
selected node. We then observed the likelihood of successful
cascades, as $p$ increases from 0 to 1, repeating the experiment
for different threshold values. The second algorithm rewires links
in such a way that nodes keep their degrees (and thus the original
degree distribution is conserved) by {\em permuting} links
\cite{Maslov03}: a link connecting nodes $i$ and $j$ is permuted
with a link connecting nodes $k$ and $l$. For both cases, a
cascade is {\em successful} if it reaches at least 90\% of network
nodes.

For $T>T_c^{2dnn}=3/8$ (the critical threshold for $p=0$),
cascades are precluded for all $p$. Permuting links such that all
nodes have the same degree $k=8$, if $T<1/8$ (the critical value
for $p=1$), cascades are guaranteed for all $p$. Thus, for
multiplex propagation randomization is only meaningful within the
window $1/8 \le T \le 3/8$. Figure~\ref{fig1} reports the phase
diagram for cascade frequency for thresholds in this range, as the
original regular neighborhoods ($\langle k \rangle=8$) are
randomized with probability $0.001 \le p \le 1$. Despite small
differences between the two algorithm used for the perturbation of
the network, the phase diagram shows that cascades are bounded
above by $T_c=3/8$ and below by $T_c=1/8$. As thresholds are
increased, the critical value of $p$ decreases, making cascades
less likely in the small-world network region of the phase space.

Figure~\ref{fig2} shows the effects of perturbation on two
neighborhoods with the focal nodes $i$ and $j$.  $i$'s
neighborhood is a seed neighborhood (shaded) and $j$'s
neighborhood (outlined) is inactive. In Figure~\ref{fig2}a, the
nodes $k$, $k'$, and $k''$ are shared by both neighborhoods $i$
and $j$. By acting as bridges between the two neighborhoods, these
nodes allow multiplex propagation to spread from one to the other.
Random rewiring reduces the width of the bridge between the
neighborhoods by reducing the common neighbors shared by $i$ and
$j$, as shown in Figure~\ref{fig2}b, where random rewiring has
eliminated two of the common neighbors of $i$ and $j$. In the
resulting network, $i$'s neighborhood can only activate $j$
through $k$; thus, if $j$ requires multiple sources of activation,
$i$'s neighborhood will no longer be sufficient to activate $j$.

When ties are randomly rewired, local changes to neighborhood
structure dramatically affects the dynamics of propagation.
Fig.~\ref{fig4} shows the different growth rates of cascades in
regular and rewired networks. In a regular lattice the growth of
active nodes follows a power law with an exponent around 2, due to
the two dimensional nature of the network. However in the small
network the growth initially follows an exponential law and then
it rapidly expands and activates all the nodes.

\begin{figure}
\epsfig{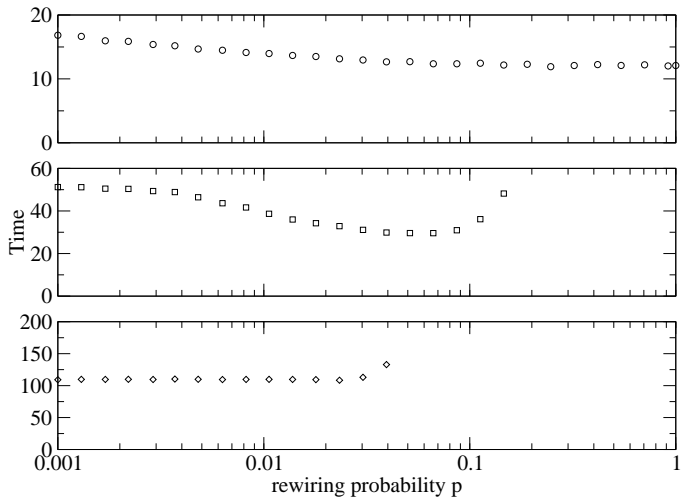} \caption{Time
to activate all the nodes ($N=10000$) for an initial seed for
$T=0.12$, $0.24$, and $0.36$. Time has been averaged over 100
realizations. The randomized networks have been obtained using the
permutation algorithm.}
\label{fig3}
\end{figure}

\begin{figure}[t]
\vspace{1cm}
\epsfig{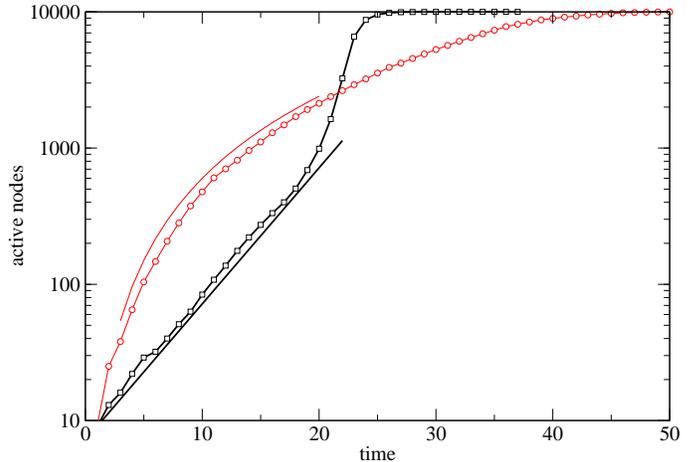}
\caption{Total number of active nodes as a function of time from
an initial seed of 9 active neighbor nodes and $T=0.24$ for one
realization. While for (circles) the regular lattice ($p=0$) the
curve follows a power law growth (solid line increases as
$(time)^2$), (squares) for a small-world network (following the
permutation algorithm with $p=0.1$) the growth is initially
exponential (solid line).}
\label{fig4}
\end{figure}

In Fig.~\ref{fig3} we show the average time required for the
initial seed to reach the full population for different values of
the threshold and the rewiring probability $p$, using the
permutation algorithm. For simple propagation, random perturbation
of ties reduces propagation time as expected. However, perturbing
the network lowers its critical threshold, thus reducing the
viability of contagions that spread by multiplex propagation.  As
shown in Fig.~\ref{fig3}, at the critical point $T_c$, rates for
multiplex propagation diverge, increasing to infinity with
increasing $p$. Thus, although random links can increase the rate
of propagation, they can also preclude propagation by lowering the
critical threshold of the network.

%%%%%%%%%%%%%%%%%%%%%%%%%%%%%%%%%%%%%%%%%%%%%%%%%%%%%%%%%%%%%%%Fig. 3

\begin{figure}[t]
\vspace{0.5cm} \epsfig{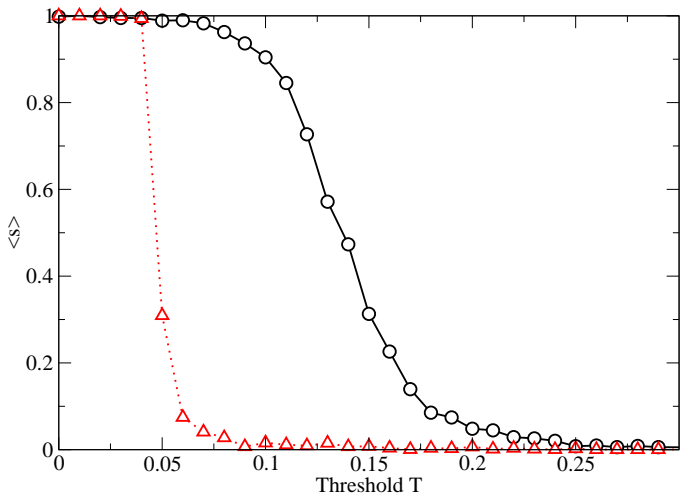}
\caption{ Effect of perturbation on multiplex propagation in the
IMDB network. For the unperturbed network (black circles), $T_c
\simeq 0.1$. For the randomized network (red triangles), $T_c
\simeq 0.04$ (approximately $1/z$ for $\langle k \rangle \simeq
25$). The perturbed network cannot support multiplex propagation
that is possible on the real structured social network. The
randomized network has been obtained permuting links in order to
keep the original degree distribution \cite{Maslov03}.}
\label{fig5}
\end{figure}

%%%%%%%%%%%%%%%%%%%%%%%%%%%%%%%%%%%%%%%%%%%%%%%%%%%%%%%%%%%%%%%%%%%%%%%%%%%%%
{\em Empirical scale-free networks.}- Regular lattices are an
important theoretical demonstration of multiplex propagation
because they can have very wide bridges between near-neighbors.
Nevertheless, except for special cases where spatial patterns of
interaction dominate the structure of the network of interaction
\cite{Gould95}, regular lattices are not a good representation of
real networks. We therefore extended our analysis to an empirical
social network. In particular, we consider the Internet Movie Data
Base as an illustrative example. Figure~\ref{fig5} reports the
average relative size of cascades, $\langle s \rangle$, in the
Internet movie database (IMDB). It is worth noting that due to the
nature of the transition for cascade behavior, the average
frequency of cascades is equivalent to the average relative size
of cascades, $\langle s \rangle$. The black line shows $\langle s
\rangle$ for the original network ($p=0$), where $T_c \simeq 0.1$,
while the red line represents the randomized IMDB ($p=1$). In the
randomized graph, $T_c \simeq 0.04$ (approximately $1/\langle k
\rangle$ for estimated $\langle k \rangle =25$). Consistent with
our results for regular lattices, the socially clustered IMDB
network supports multiplex propagation that cannot propagate on
the randomized network.

%%%%%%%%%%%%%%%%%%%%%%%%%%%%%%%%%%%%%%%%%%%%%%%%%%%%%%%%%%%%%%%Fig. 4

%%%%%%%%%%%%%%%%%%%%%%%%%%%%%%%%%%%%%%%%%%%%%%%%%%%%%%%%%%%%%%%%%%%%%%%%%%%%%
{\em Conclusions.}-- Using a threshold model, we have analyzed
simple and multiplex propagation in different classes of complex
networks. The relevant bridging mechanism for multiplex
propagation is not the dyadic link but multiple short paths
between source and target. As a regular lattice is randomized,
there are fewer common neighbors to provide multiple simultaneous
sources of activation. Thus, while networks with long range links
have been shown to promote simple propagation in small-world
networks, they inhibit multiplex propagation. This implies that
random links do not promote diffusion if the credibility of
information or the willingness to adopt an innovation depends on
receiving independent confirmation from multiple sources.

%%%%%%%%%%%%%%%%%%%%%%%%%%%%%%%%%%%%%%%%%%%%%%%%%%%%%%%%%%%%%%%Fig. 5

The qualitative differences between multiplex and simple
propagation caution about extrapolating from the spread of disease
or information to the spread of participation in political,
religious, or cultural movements. These movements may not benefit
from ``the strength of weak links" and may even be hampered by
processes of global integration. More broadly, many of the
important empirical studies of the effects of small-world networks
on the dynamics of cascades may need to take into account the
possibility that propagation may be multiplex. In fact, for the
dynamics of multiplex propagation, our results highlight the
inhibitory effects of networks typically thought to be
advantageous for cascades.

%%%%%%%%%%%%%%%%%%%%%%%%%%%%%%%%%%%%%%%%%%%%%%%%%%%%%%%%%%%%%%%%%%%%%%%%%%%%%
D.M.C. acknowledges support from a NSF IGERT Fellowship and a NSF
Human Social Dynamics grant (SES-0432917). M.W.M. acknowledges
support from the NSF (SES-0241657 and SES-0432917). V.M.E
acknowledges support from McyT (Spain) through project CONOCE2. We
also thank Steve Strogatz, Duncan Watts and Jon Kleinberg for
helpful comments and suggestions.

%%%%%%%%%%%%%%%%%%%%%%%%%%%%%%%%%%%%%%%%%%%%%%%%%%%%%%%%%%%%%%%%%%%%%%%%

%%%%%%%%%%%%%%%%%%%%%%%%%%%%%%%%%%%%%%%%%%%%%%%%%%%%%
\end{document}